\documentclass[iop]{emulateapj}

\usepackage{natbib,aas_macros}
\usepackage{multirow,color}

\begin{document}
\shorttitle{
Lensing Interpretation of Ultra-Luminous QSO %at $z=6.30$
}
\shortauthors{Fujimoto et al.}
\slugcomment{ApJ in press}

\title{%
Truth or delusion? A possible gravitational lensing interpretation of \\
the ultra-luminous quasar %J0100+2802
SDSS J010013.02+280225.8 at $\lowercase{z}=6.30$
}

\author{%
Seiji Fujimoto\altaffilmark{1,2,3,4,5}, 
Masamune Oguri \altaffilmark{6,7,8},
Tohru Nagao\altaffilmark{9},
Takuma Izumi\altaffilmark{4,10},
and Masami Ouchi \altaffilmark{4,5,8}
}

\email{fujimoto@nbi.ku.dk}

\altaffiltext{1}{%
Cosmic DAWN Center 
}
\altaffiltext{2}{%
Niels Bohr Institute, University of Copenhagen, Lyngbyvej 2, DK-2100, Copenhagen, Denmark
}
\altaffiltext{3}{%
Research Institute for Science and Engineering, Waseda University, 3-4-1 Okubo, Shinjuku, Tokyo 169-8555, Japan
}
\altaffiltext{4}{%
National Astronomical Observatory of Japan, 2-21-1, Osawa, Mitaka, Tokyo, Japan
}
\altaffiltext{5}{%
Institute for Cosmic Ray Research, The University of Tokyo,
Kashiwa, Chiba 277-8582, Japan
}
\altaffiltext{6}{%
Research Center for the Early Universe, University of Tokyo, 7-3-1 Hongo, Bunkyo-ku, Tokyo 113-0033, Japan
}
\altaffiltext{7}{%
Department of Physics, University of Tokyo, 7-3-1 Hongo, Bunkyo-ku, Tokyo 113-0033, Japan
}
\altaffiltext{8}{%
Kavli Institute for the Physics and Mathematics of the Universe (Kavli IPMU, WPI), University of Tokyo, Chiba 277-8583, Japan
}
\altaffiltext{9}{%
Research Center for Space and Cosmic Evolution, Ehime University, 2-5 Bunkyo-cho, Matsuyama, Ehime 790-8577, Japan
}
\altaffiltext{10}{%
Department of Astronomical Science, Graduate University for Advanced Studies (SOKENDAI), 2-21-1 Osawa, Mitaka, Tokyo 181-8588, Japan
}

\newcommand{\oiii}{[O\,{\sc iii}]}
\newcommand{\oii}{[O\,{\sc ii}]}
\newcommand{\cii}{[C\,{\sc ii}]}
\newcommand{\ciii}{C\,{\sc iii}]}
\newcommand{\lya}{Ly$\alpha$}
\newcommand{\mum}{$\mu$m}
\newcommand{\dv}{$\Delta v_{\rm Ly\alpha}$}
\newcommand{\ew}{EW$_{\rm 0}$}
\newcommand{\lsun}{$L_{\rm \odot}$}
\newcommand{\msun}{$M_{\rm \odot}$}
\newcommand{\ltir}{$L_{\rm TIR}$}
\newcommand{\nhi}{$N_{\rm HI}$}
\newcommand{\loiii}{$L_{\rm [OIII]}$}
\newcommand{\llya}{$L_{\rm Ly\alpha}$}
\newcommand{\luv}{$L_{\rm UV}$}
\newcommand{\zph}{$z_{\rm ph}$}
\newcommand{\muv}{$M_{\rm UV}$}
\newcommand{\td}{$T_{\rm d}$}
\newcommand{\bd}{$\beta_{\rm d}$}
\newcommand{\md}{$M_{\rm d}$}
\newcommand{\zoiii}{$z_{\rm [OIII]}$}
\newcommand{\zlya}{$z_{\rm Ly\alpha}$}
\newcommand{\zsun}{$Z_{\rm \odot}$}
\newcommand{\mdyn}{$M_{\rm dyn}$}

\def\apj{ApJ}%
         % Astrophysical Journal
\def\apjl{ApJ}%
         % Astrophysical Journal, Letters
\def\apjs{ApJS}% 
         % Astrophysical Journal, Supplement

\def\rme{\rm e}
\def\rmFIR{\rm FIR}
\def\itHubble{\it Hubble}
\def\rmyr{\rm yr}

%---------------------------------------------------------------------
\begin{abstract}
Gravitational lensing sometimes dominates the observed properties of apparently very bright objects.
We present morphological properties in the high-resolution (FWHM $\sim0\farcs15$) Atacama Large Millimeter/submillimeter Array (ALMA) 1-mm map for an ultra-luminous quasar (QSO) at $z=6.30$, SDSS J010013.02+280225.8 (hereafter J0100+2802), whose black hole mass $M_{\rm BH}$ is the most massive ($\sim$ 1.2$\times10^{10}M_{\odot}$) at $z>6$ ever known. 
We find that the continuum emission of J0100+2802 is resolved into a quadruple system within a radius of $0\farcs2$, 
which can be interpreted as either multiple dusty star-forming regions in the host galaxy or multiple images due to strong gravitational lensing. 
The Mg {\sc ii} absorption and the potential Ly$\alpha$ line features have been identified at $z=2.33$ in the near-infrared spectroscopy towards J0100+2802, and a simple mass model fitting well reproduces the positions and flux densities of the quadruple system, both of  which are consistent with the latter interpretation. 
Although a high-resolution map taken in the Advanced Camera for Survey (ACS) on board {\it Hubble Space Telescope} (HST) shows a morphology with an apparently single component, in our fiducial lens mass model it can simply be explained by a $\sim50$ pc scale offset between the ALMA and HST emission regions.
In this case, the magnification factor for the observed HST emission is obtained to $\sim450$, reducing the intrinsic $M_{\rm BH}$ estimate to even below $10^{9}\,M_{\odot}$. 
The confirmation or the rejection of the gravitational lensing scenario is important for 
our understanding of the super-massive black holes in the early Universe.  
\end{abstract}

%---------------------------------------------------------------------
%---------------------------------------------------------------------

%%%%%%%%%%%%%%%%%%%%%%%%%%%%%%%%%%%%%%%%%%%%%%%%%%%%%%%%%%%%%%%%%
\section{Introduction}
\label{sec:intro}

The existence of the super-massive black hole (SMBH) in the early Universe \citep[e.g.,][]{wu2015,banados2018} challenges to the theories \citep[e.g.,][]{volonteri2006} of the formation and growth of the black holes (BHs). 
Luminous quasars (QSOs) at high redshift are massive galaxies hosting the SMBHs in their centers and thus serve a unique laboratory to study the evolution mechanism of the SMBH as well as the earliest phase of galaxy formation and evolution. 

In the last two decades, more than 100 $z\sim6$ QSOs have been discovered through wide-field surveys in the optical--near infrared (NIR) wavelengths \citep[e.g.,][]{jiang2009,jiang2016,willott2010,venemans2013,venemans2015,banados2016,matsuoka2016,matsuoka2018b}. 
Currently, the most massive BH at $z>6$ ever known is J0100+2802 at $z=6.30$ \citep{wu2015}, 
originally identified owing to its red optical color with the dataset of the Sloan Digital Sky Survey (SDSS; \citealt{york2000}), 
two Micron All Sky Survey (2MASS; \citealt{skrutskie2006}) and the Wide-field Infrared Survey Explorer (WISE; \citealt{wright2010}).
The bolometric luminosity $L_{\rm bol}$ is estimated to be 
$4.29\times10^{14}\,L_{\odot}$ based on an empirical conversion factor from the luminosity at 3,000 ${\rm \AA}$,  
while the BH mass $M_{\rm BH}$ is evaluated to be 1.24$\,\pm\,0.19\times10^{10}\,M_{\odot}$ via the single-epoch virial $M_{\rm BH}$ estimator based on the Mg{\sc ii} line \citep[e.g.,][]{vestergaard2009}. 
Although the ten billion solar mass BH at $z=6.30$ is reproduced under the assumptions of the Eddington-limited accreting rate and the BH seed mass of at least 1,000 $M_{\odot}$ by $z=40$, 
it is yet to be known whether these assumptions are valid.
Moreover, it is also an open question how to lose the angular momentum in the inter-stellar medium ($>$ 100 pc) 
and keep the Eddington-limited mass transportation to the accretion disk ($<$ 1 pc) around the central BH \citep[e.g.,][]{sugimura2018} under the strong feedback effect \citep[e.g.,][]{park2017,latif2018}.

The Atacama Large Millimeter / submillimeter Array (ALMA) enables us to investigate the star-forming properties of the SMBH host galaxies at $z\sim6$ via the far-infrared fine-structure lines, such as \cii\ 158 $\mu$m and \oiii\ 88 $\mu$m, and the dust continuum emission \citep[e.g.,][]{wang2013,decarli2018,venemans2018, walter2018, hashimoto2018c}. 
These ALMA studies reveal that the SMBH host galaxies at $z\sim6$ have intense star-formation rates of $\sim$ 100 -- 3000 $M_{\odot}$ yr$^{-1}$.  
The intense star formation may cause the turbulence and contribute to the angular momentum dissipation in the inter-stellar medium \citep[e.g.,][]{kawakatu2009, izumi2016}. 
The size, structure, and dynamics of these star-forming regions in the SMBH host galaxies are thus helpful probes to investigate the mass accretion to the SMBH under the star-forming activities of their host galaxies. 
However, the majority of these ALMA studies do not resolve the host galaxy structure due to the moderate angular resolution of $\sim1\farcs0$. 

In this paper, we investigate the detailed structure of the host galaxy of J0100+2802 with the high-resolution ($\sim0\farcs15$) ALMA Band 6 data. 
Since J0100+2802 is known to be the SMBH in the early Universe as the most massive BH
ever known at $z>6$, 
this is an essential step to understand the spatially-resolved star-forming nature around the early SMBHs. 
The structure of this paper is as follows. 
The ALMA observations and the data reduction are described in Section 2. 
Section 3 outlines the data analysis and interpretation. 
In Section 4, we discuss the physical properties of J0100+2802. 
A summary of this study is presented in Section 5. 

%%%%%%%%%%%%%%%%%%%%%%%
\begin{figure*}[t]
\begin{center}
\includegraphics[trim=0cm 0cm 0cm 0cm, clip, angle=0,width=1.0\textwidth]{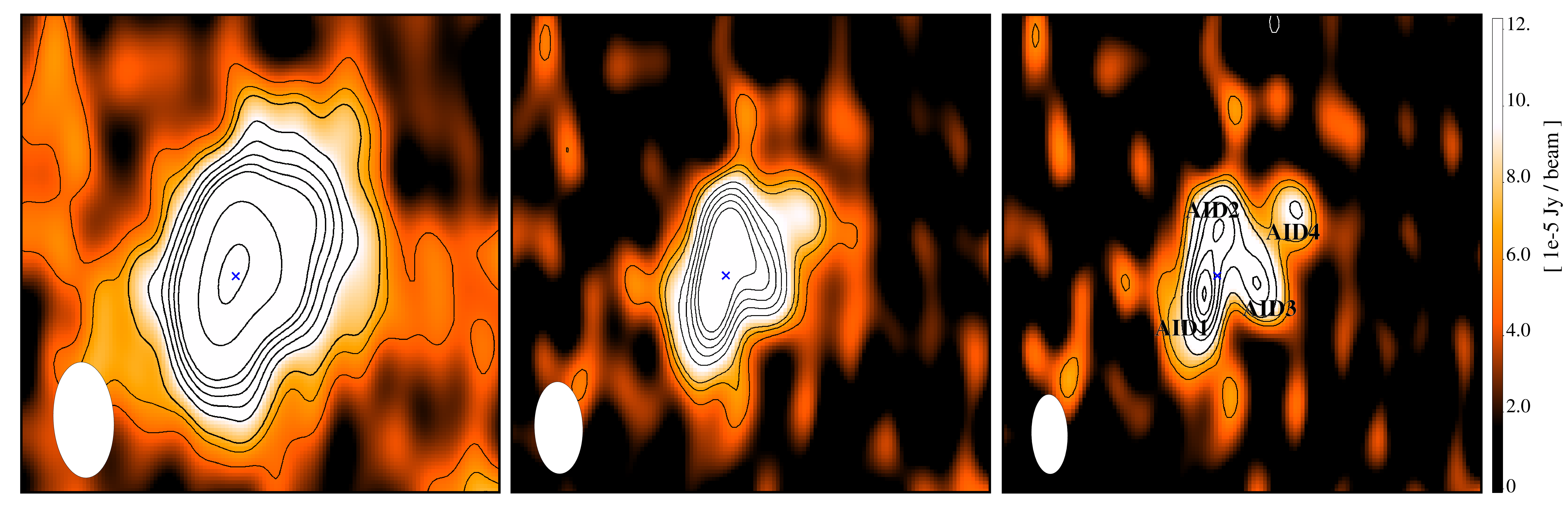}
 \caption[]{
ALMA $1\farcs2\times1\farcs2$ images of J0100+2802. 
The LR (natural-weighted), MR (briggs-weighted, robust = 0.5), and HR (briggs-weighted, robust = 0.2) maps are presented from left to right. 
The white contour shows the $-3\sigma$ level, 
and the black contour denotes the 3$\sigma$, 4$\sigma$, 5$\sigma$, 6$\sigma$, 7$\sigma$, 8$\sigma$, 9$\sigma$, 10$\sigma$, 15$\sigma$, and 20$\sigma$ levels. 
The rms noise levels of the LR, MR, and HR maps are 16 $\mu$Jy/beam, 17 $\mu$Jy/beam, and 20 $\mu$Jy/beam, respectively.
The ALMA synthesized beam is presented at the bottom left. 
We confirm that the MR map shows the consistent morphology in the previous study (see Figure 1 in \citealt{wang2019}). 
The blue cross represents the optical emission peak position in the GAIA DR2 catalog. 
\label{fig:alma_maps}}
\end{center}
\end{figure*}
%%%%%%%%%%%%%%%%%%%%%%%

Throughout this paper, we assume a flat universe with 
$\Omega_{\rm m} = 0.3$, 
$\Omega_\Lambda = 0.7$, 
$\sigma_8 = 0.8$, 
and $H_0 = 70$ km s$^{-1}$ Mpc$^{-1}$. 
We use magnitudes in the AB system \citep{oke1983}.

%%%%%%%%%%%%%%%%%%%%%%%%%%%%%%%%%%%%%%%%%%%%%%%%%%%%%%%%%%%%%%%%%
\section{Data \& Reduction} 
\label{sec:data}

Observations were carried out on 2016 September 4 in the cycle 4 program (PI: X. Fan; see \citealt{wang2019}), using 44 antennas with the projected baselines ranging from 15 m to 2.5 km. 
The available 7.5 GHz bandwidth with four spectral windows was centered at an observed frequency of 239.8 GHz (i.e., $\sim$1.25 mm).
J2253+1608 and J0238+1636 were observed as the flux calibrator, while J2253+1608 and J0237+2848 were used for the bandpass calibrator. 
Phase calibration was generally performed by using observations of J0057+3021.
The total on-source time was $\sim$ 74 min. 

We reduce the ALMA data with the Common Astronomy Software Applications package version 4.7.0 (CASA; \citealt{mcmullin2007}) in the standard manner with the scripts provided by the ALMA observatory. 
For further data analyses, we use the CASA version 5.4.0. 
The continuum images are produced with the line-free channels in all spectral windows
by the CLEAN algorithm with the {\it tclean} task. 
The CLEAN boxes are set at the peak pixel positions with S/N $\geq$ 5 in the auto mode, 
and the CLEAN routines are proceeded down to the 3$\sigma$ level. 
The final natural-weighted image is characterized by a synthesized beam size of $0\farcs29\times0\farcs15$ and the rms noise level of 16 $\mu$Jy/beam.  
We also produced a briggs-weighted maps with the robust parameter of 0.5 and 0.2 whose final synthesized beam sizes are $0\farcs23\times0\farcs12$ and $0\farcs21\times0\farcs09$ with the rms noise levels of 17 $\mu$Jy/beam and 20 $\mu$Jy/beam, respectively. 
We refer to the natural-, the briggs-  (robust = 0.5), the briggs-weighted  (robust = 0.2) maps as ``LR'', ``MR'', and ``HR'' maps, respectively. 
Although the uniform-weighting (robust $=-2.0$) produces a higher-resolution map, the noise level increases. 
Here we do not use the uniform-weighted map to obtain reliable results with secure signal-to-noise ratios (SNRs).  
Note that the imaging parameter of the MR map is the same as the continuum map produced in \cite{wang2019}, 
and we confirm that the same morphology is reproduced in the MR map as the previous works (see Figure 1 in \citealt{wang2019}). 

the continuum map was produced by utilizing the line-free channels in all spectral windows,

\section{Analysis \& Interpretation}
\label{sec:analysis}

\subsection{ALMA}
\label{sec:alma}

In Figure \ref{fig:alma_maps}, we show the LR, MR, and HR maps of J0100+2802.  
In the LR map, a continuum emission is clearly detected with the 21$\sigma$ level at the peak. 
The total flux density is estimated to be 1.3 mJy with an aperture diameter of $0\farcs8$, 
which is consistent with the estimate in \cite{wang2019}. 
The position at the peak and the total flux density are summarized in Table \ref{tab:comp_summary}. 

%ttttttttttttttttttttttttttttttttttttttttttttt
\begin{table}
\begin{center}
{%\scriptsize
\caption{Component Summary
\label{tab:comp_summary}}
\begin{tabular}{cccccc}
\hline
\hline
Comp. & R.A.          & Decl.         & $S_{\rm obs}$   & SNR    &  \\
      &  (J2000)      &  (J2000)     &  (mJy)          &        &  \\
  (1) &\multicolumn{2}{c}{(2)}        &   (3)           &  (4)   &  \\ \hline
Total & 1:00:13.026    & +28:02:25.81     & 1.3   & 21   &  \\ \hline
      &  $\Delta\alpha$ &  $\Delta\delta$   &  $S_{\rm obs, peak}$     &      &  $\mu$  \\ 
      &    ($''$)       &   ($''$)          &  (mJy/beam)     &      &        \\ 
      &     (5)         &    (6)            &    (7)   &      &   (8)  \\ \hline      
AID1   &   +0.02    & $-$0.05  & 0.19  & 10    & 19.6 \\
AID2   &  $-$0.01  &  +0.12    & 0.17  & 8.9   & 15.2 \\
AID3   &  $-$0.13  & $-$0.02   & 0.14  & 7.4   & 13.9 \\
AID4   &  $-$0.22  & +0.16     & 0.10  & 5.4   & 9.7 \\ \hline
\end{tabular}
}
\end{center}
\tablecomments{
\footnotesize{
(1) Name of the components. 
(2) Coordinate of the 1-mm continuum emission evaluated with the peak pixel position. 
(3) Flux density estimated with the aperture diameter of $0\farcs8$. 
(4) Signal-to-noise ratio at the peak. 
(5) R.A. offset from (1). 
(6) Decl. offset from (1). 
(7) Peak pixel value for each component. 
(8) Magnification factor at the peak estimated with the fiducial mass model. 
}}
\end{table}

%%%%%%%%%%%%%%%%%%%%%%%%%%%%%%%%%%%%%%%%%%%%%%%%%%%%%%%%%%%%%%%%%

In the HR map, on the other hand, the continuum emission is resolved into multiple peaks. 
We identify four peaks with the positive counts above the 5.0$\sigma$ level that are located within a radius of $\sim0\farcs2$, 
while no negative peaks are detected below the $-4.0\sigma$ level in the HR map. 
We refer to these four components from bright to faint as AID1, AID2, AID3, and AID4. 
In Table \ref{tab:comp_summary}, 
we summarize the peak counts, SNRs, and the positions for these four components. 
The sum of the peak counts of the four components reaches 0.6 mJy that corresponds to $\sim46\%$ ($=0.6/1.3$) of the total flux density. 
This suggests that about half of the continuum is the resolved, diffused emission in the HR map. 
Note that there is a possibility that the combination of the diffuse continuum and the noise fluctuation 
causes even over the 5.0$\sigma$ positive peaks \citep[e.g.,][]{hodge2016}.
To evaluate this probability, we have performed a mock observation with the CASA task {\sc simobserve} towards the diffuse continuum. 
Here we assume a uniform flux distribution whose total flux is estimated from the resolved, diffuse emission of 0.7 mJy (= 1.3 mJy -- 0.6 mJy) with the diameter of $0\farcs8$. 
We then obtain the visibility data set through {\sc simobserve} and produce the HR map. 
We repeat from the mock observation to producing the HR map 1,000 times, and identify positive peaks within the radius of $0\farcs2$ from the center of the diffuse emission. 
We find that 8 out of the 1,000 HR maps contain the positive peaks over the 5.4$\sigma$ level that are likely to be produced by the diffuse continuum and the noise fluctuation. 
No positive peak over the 7$\sigma$ level is detected. 
These results indicate that AID4 (SNR $=$ 5.4) may be caused by the combination of the diffuse continuum and the noise fluctuation with the probability of $\sim$ 0.8\%, 
while AID1, AID2, and AID3 (SNR $>$ 7) are hard to be explained by this combination. 
We also calculate the surface brightness of the diffuse continuum at the AID4 position in other profiles (e.g., Gaussian, exponential-disk, deVaucouleurs-spheroidal) with the effective radii of 0.5 -- 2.0 kpc which are the typical values among $z\sim6$ star-forming galaxies with the stellar mass $M_{\rm star}$ of $\sim10^{10.5-11}$ M$_{\odot}$ \citep[e.g.,][]{shibuya2015}. 
We find that the surface brightness values become fainter than one in the uniform profile, 
indicating that AID4 is still hard to be explained by modeling of the different surface brightness profiles.

\begin{figure}
\begin{center}
\includegraphics[trim=0cm 0cm 0cm 0cm, clip, angle=0,width=0.5\textwidth]{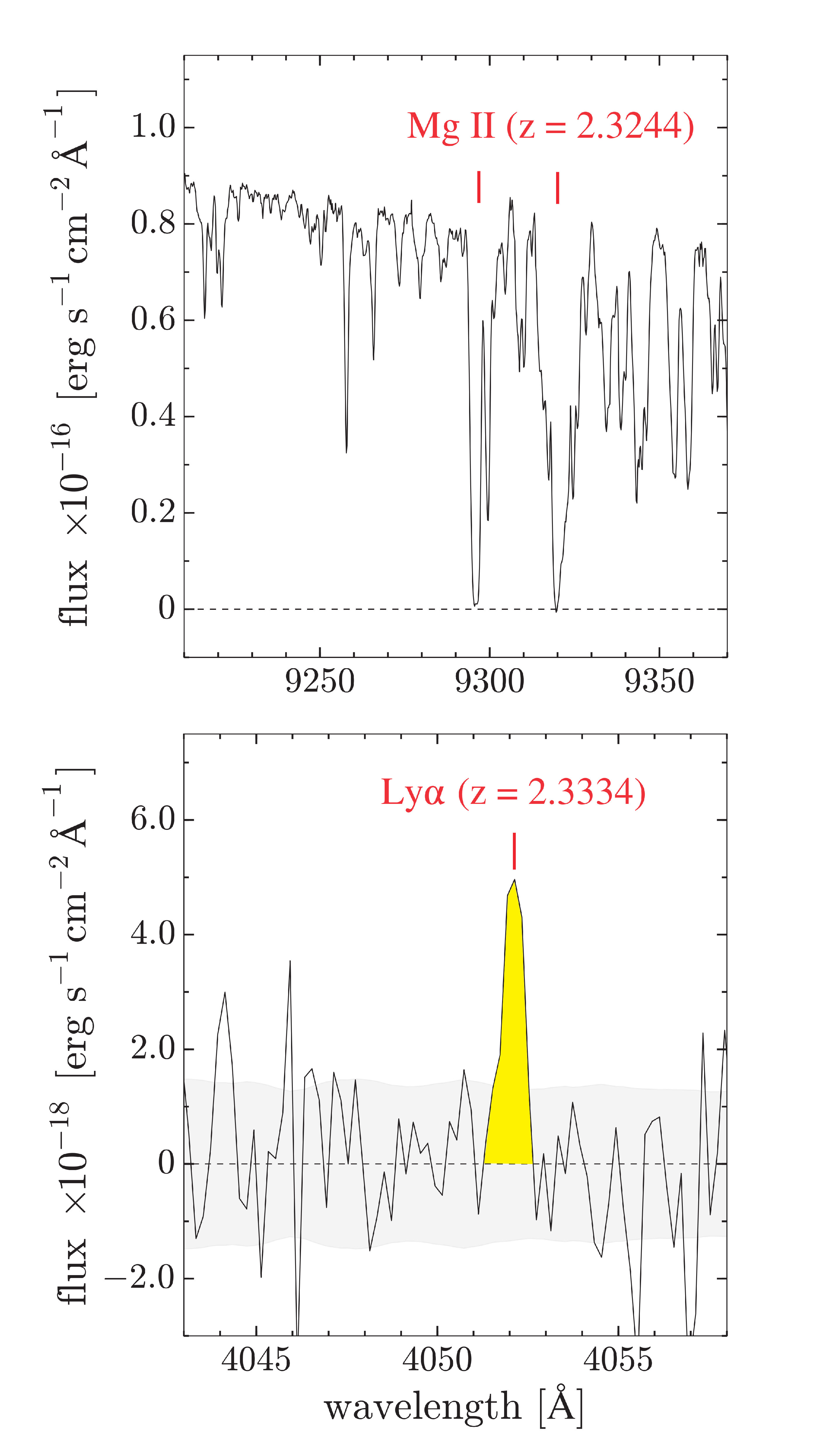}
 \caption[]{
The optical-NIR spectroscopy towards J0100+2802 taken with VLT/X-shooter. 
{\bf Top:}
The Mg {\sc ii} absorption doublet is identified at $z=2.3244 \pm 0.0002$ which is consistent with the previous report in \cite{wu2015}.  
{\bf Bottom:}
Possible Ly$\alpha$ line emission at $z=2.3334 \pm 0.0002$ with the $5.3\sigma$ level. 
The gray shaded area represents the 1$\sigma$ flux uncertainty. 
Both spectra are produced by weighted-average of individual processed spectra on the ESO archive (096.A-0095; PI: M. Pettini). 
The redshifts of the Mg {\sc ii} absorption and the Ly$\alpha$ line are estimated from the peak wavelengths, where the uncertainty is calculated from the spectral resolution.  
\label{fig:absorption}}
\end{center}
\end{figure}

One interpretation for these four components is that they are quadruple images due to strong gravitational lensing effect \citep[e.g.,][]{magain1988}. 
In fact, 
\cite{wu2015} report the existence of abundant absorbers (e.g., at $z=2.33$ and 3.34) in the optical--NIR spectroscopy towards J0100+2802. 
In the top panel of Figure \ref{fig:absorption}, we show the Mg {\sc ii} absorption line feature at $z=2.3244\pm0.0002$ which we confirm in the latest optical--NIR spectroscopy with VLT/X-shooter from the ESO archive (PI: M. Pettini; see \citealt{becker2019}). 
Moreover, the bottom panel of Figure \ref{fig:absorption} shows the possible Ly$\alpha$ line emission at $z=2.3334\pm0.0002$ with the $5.3\sigma$ level 
that we also identify in the same data from VLT/X-shooter. 
The Ly$\alpha$ velocity offset from the Mg {\sc ii} absorption is estimated to be $810\pm90$ km s$^{-1}$ which falls in the general range of the Ly$\alpha$ velocity offset among $z\sim2$ galaxies (see e.g., Figure 8 in \citealt{stark2017}). 
We obtain the luminosity of the potential Ly$\alpha$ line $L_{Ly\alpha} = 1.1\times10^{42}$ erg s$^{-1}$. 
This can be converted into the star-formation (SFR) value of $\sim1$ M$_{\odot}$ yr$^{-1}$ with the scaling relation of SFR $=9.1\times10^{-43}\times L_{Ly\alpha}$ M$_{\odot}$ yr$^{-1}$ \citep{kennicutt1998}. 
Based on the typical values between SFR and $M_{\rm star}$ among quiescent galaxies at the similar redshift, 
the SFR value of $\sim1$ M$_{\odot}$ yr$^{-1}$ corresponds to the $M_{\rm star}$ value of $\sim10^{10-11}$ M$_{\odot}$ (see e.g., Fig. 3 in \citealt{barro2014}). 
These results suggest the possibility that the foreground object with $M_{\rm star}\sim10^{10-11}$ M$_{\odot}$ at $z=2.33$ may affect the brightness of J0100+2802 via gravitational lensing effect.

%%%%%%%%%%%%%%%%%%%%%%%
%%%%%%%%%%%%%%%%%%%%%%% 

To test whether the four components identified in the HR map can indeed be explained by 
strong grtavitational lensing,
we construct a mass model with the parametric gravitational lensing package {\sc glafic} \citep{oguri2010}. Here we fix the lens redshift at $z=2.33$ where the possible foreground object is identified.
The mass model consists of a singular isothermal ellipsoid (SIE) and an external shear. 
We adopt no priors on the centroid of the SIE, 
while we add a Gaussian prior on the amplitude of the external shear of $\gamma=0.05 \pm 0.05$. 
The flux errors are assumed to be 10 \%. 
The approximate positional uncertainty of the ALMA map $\Delta p$ in milliarcsec is given by 
\begin{eqnarray}
\Delta p = \frac{70000}{\nu * B * \sigma},
\end{eqnarray}
where $\sigma$ is the peak SNR, $\nu$ is the observing frequency in GHz, and $B$ is the maximum baseline length in kilometers (ALMA technical handbook \footnote{Section 10.5.2: https://almascience.nao.ac.jp/documents-and-tools/cycle7/alma-technical-handbook}). 
For a peak with SNR = 5 in our ALMA maps, we obtain $\Delta p $ of $\sim0\farcs02$.  
Due to the large elongation of the ALMA beam shape, we conservatively adopt the positional error of $0\farcs03$. 
After the fitting routine for the four peak positions in the HR map, 
we obtain the best-fit mass model with the $\chi^2$ over the degree of freedom of 6.11/3. 
With this fiducial lens mass model, the Einstein radius is 0\farcs14 which corresponds to 
the line-of-sight velocity dispersion of the lensing galaxy of 121~km\,s$^{-1}$ at the lens redshift of 2.33. 
Interestingly, 
from the scaling relation for quiescent galaxies \citep[e.g.,][]{zahid2016}, the velocity dispersion of 121~km\,s$^{-1}$ corresponds to the $M_{\rm star}$ value of $\sim10^{10.3}$ M$_{\odot}$ which is consistent with the independent $M_{\rm star}$ estimate of $\sim10^{10-11}$ M$_{\odot}$  from the Ly$\alpha$ line luminosity. %$L_{Ly\alpha}$. 
We list the magnification factors for the four components in Table \ref{tab:comp_summary}.  The best-fitting mass model predicts the total magnification of $\sim 58$ for the ALMA source, although given the relatively large positional uncertainty the uncertainty of the lens mass model is also relatively large, with the total magnification range of $\sim 18-117$ at 2$\sigma$ level. Since the purpose of this paper is to present a possibility of the gravitational lensing interpretation rather than the full exploration of the lens mass model, in what follows we focus only on our best-fitting mass model and adopt it as our fidcuial model.

%%%%%%%%%%%%%%%%%%%%%%%
\begin{figure*}
\begin{center}
\includegraphics[trim=0cm 0cm 0cm 0cm, clip, angle=0,width=0.9\textwidth]{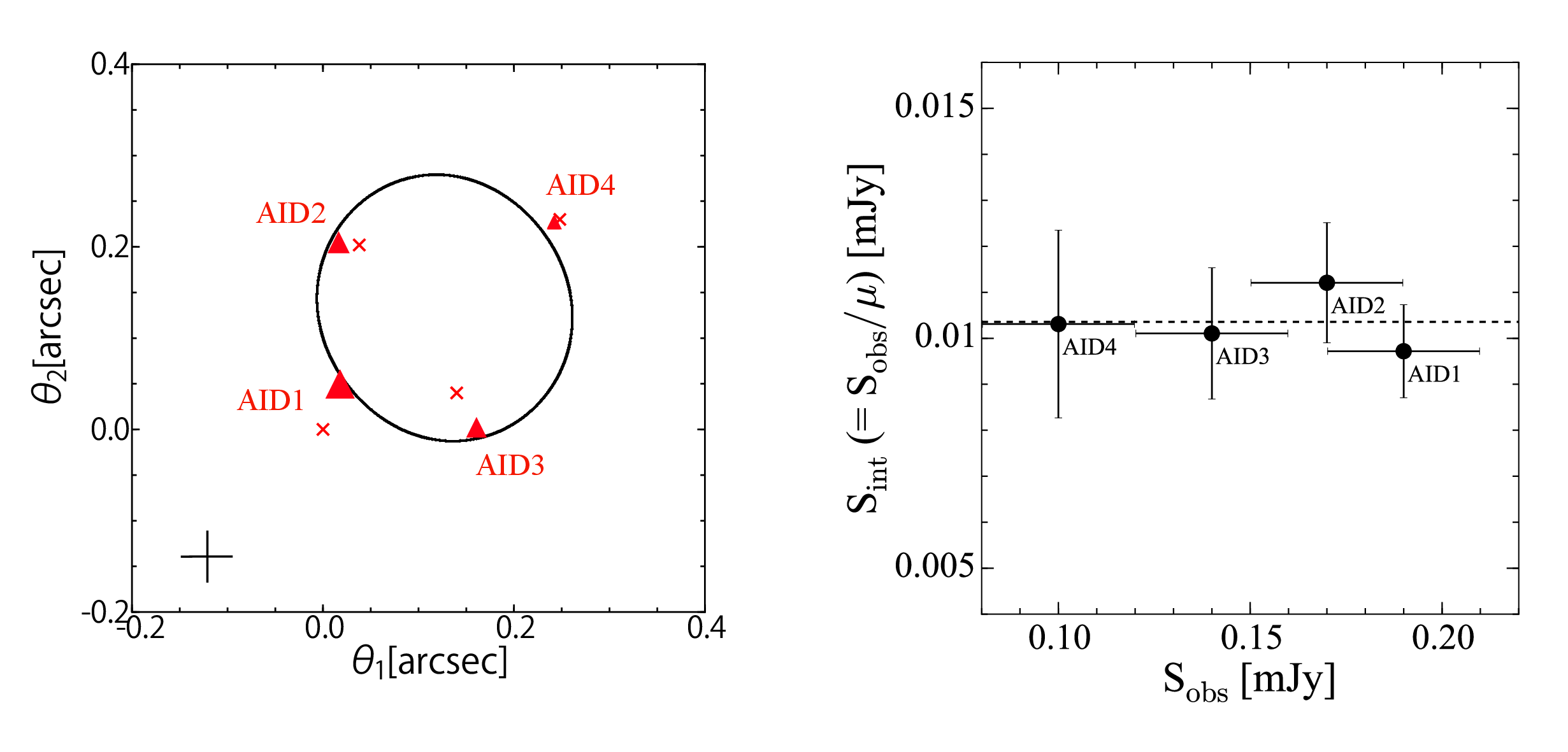}
 \caption[]{
Simple mass model fitting results for positions and flux densities of the four components identified in the ALMA HR map. 
{\bf Left:} 
The red crosses and triangles show the peak positions of the four components in the observed and the model maps, respectively.
The size difference among the red triangles correspond to the ratio of the magnification factors of the four components in our model.   
The black curve denotes the critical curve.
The error scale used in the fitting is presented at the bottom left. 
The center of the coordinate is defined at the position of AID1. 
{\bf Right:} 
The observed ($S_{\rm obs}$) and intrinsic ($S_{\rm int}$) flux densities of the four components.   
The dashed line presents the average value of the intrinsic flux densities for the four components.
\label{fig:model}}
\end{center}
\end{figure*}
%%%%%%%%%%%%%%%%%%%%%%% 

Figure \ref{fig:model} presents the positions (left panel) and intrinsic flux densities (right panel) of the four components predicted by our fiducial mass model. 
In the left panel, the red crosses and triangles show the positions of the four components in the observed and the model maps, respectively.  
We find that all the four positions are consistent within the errors. 
In the right panel, the dashed line presents the average value of the intrinsic flux densities for the four components.
We find that the intrinsic flux densities of all four components agree with the average value within the errors. 
These results quantitatively support that the four components can be explained by 
multiple images due to strong gravitational lensing.

%%%%%%%%%%%%%%%%%%%%%%%
\begin{figure*}
\begin{center}
\includegraphics[trim=0cm 0cm 0cm 0cm, clip, angle=0,width=1.0\textwidth]{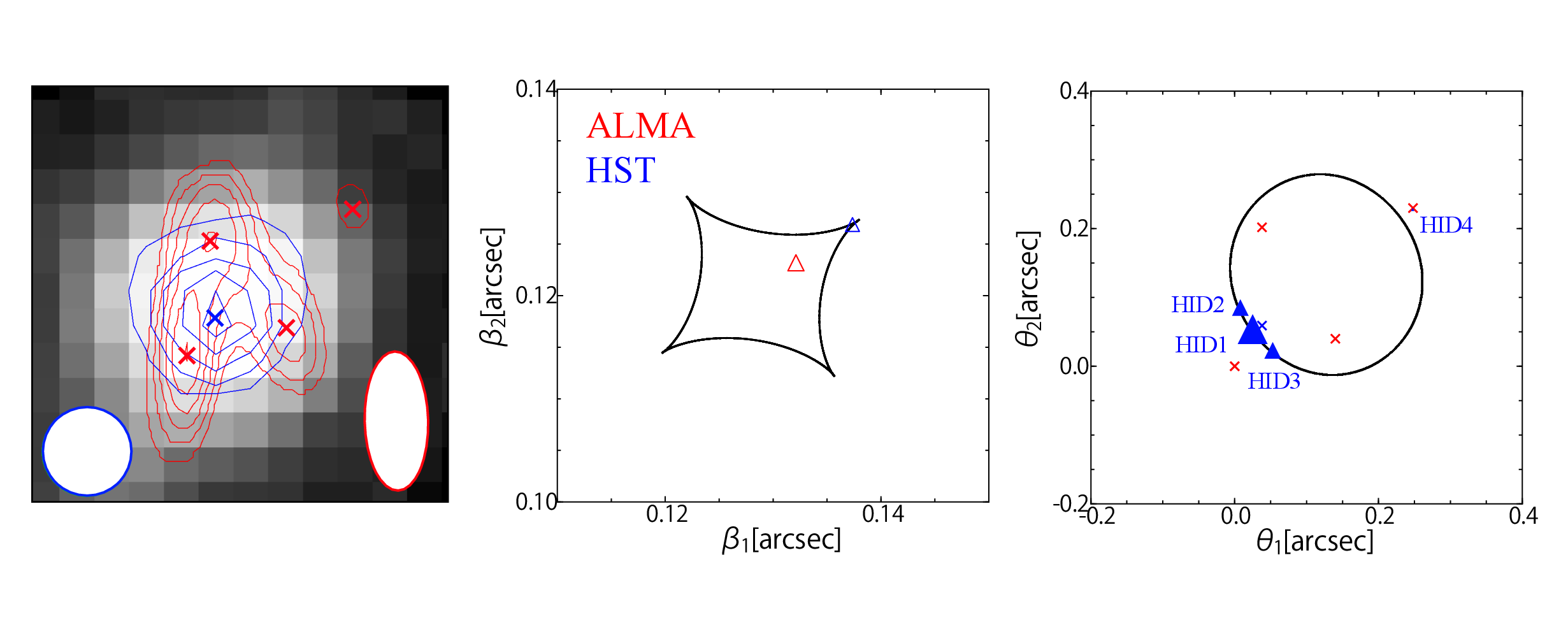}
 \caption[]{
{\bf Left: }The HST/F850LP $0.''6\times0.''6$ image for J0100+2802. 
The blue and red contours denote the continuum emission identified in the HST/F850LP and the ALMA HR maps, respectively. 
The crosses indicate the peak positions of these emission. 
The PSFs of the HST and ALMA maps are presented in the bottom left and right, respectively. 
{\bf Middle: }
The continuum peak positions in the source plane. 
The red open triangle indicates the best-fit position of the ALMA continuum emission in the simple mass model fitting (Section \ref{sec:alma}), 
and the blue open triangle shows a possible position of the HST continuum emission. 
The black line is the caustics. 
{\bf Right: }
The continuum peak positions in the image plane. 
The red and blue crosses are assigned in the same manner as the left panel. 
The blue triangles denote a possible peak positions of the HST emission in the resolution-free map predicted with our fiducial mass model, 
if the HST emission originates near from the cusp of the caustic in the source plane.  
The size difference among the triangles correspond to the ratio of the magnification factors of the four components in our fiducial mass model.   
The black line denotes the critical curve. 
In the middle and right panels, the same coordinate system is assigned as the left panel of Figure \ref{fig:model}. 
\label{fig:hst_crit_caus}}
\end{center}
\end{figure*}
%%%%%%%%%%%%%%%%%%%%%%% 

\subsection{Comparison with HST}
\label{sec:hst}

We compare the morphology in the ALMA HR map with a high-resolution data of the {\it Hubble Space Telescope} (HST) Advanced Camera for Surveys (ACS) in F850LP (PSF$\sim0.''1$). 
In the left panel of Figure \ref{fig:hst_crit_caus}, 
we present the HST/F850LP map for J0100+2802 obtained from the final flat-field and flux calibrated science products in the Hubble Legacy Archive (PI: X. Fan). 
For comparison, we also show the contours of the ALMA emission in the HR map. 
The astrometry between HST and ALMA is corrected based on the GAIA DR2 catalog \citep{gaia2018}. 
Interestingly, we find that the morphology of the HST emission 
appears to consist of a single component, in marked contrast to the ALMA emission. 
Since the HST and ALMA emissions trace the accretion disk around the central SMBH and the dusty star-forming region in the host galaxy, respectively, 
the following two possibilities arise to explain the different morphologies of J0100+2802 in the HST and ALMA maps: 
1) an offset between the HST and ALMA emission regions in the 
source plane causes the difference in gravitational lensing effects and produces the different morphologies in the image plane. 
2) the four components in the ALMA map represent multiple dusty star-forming regions in the host galaxy or a dusty on-going merging system, 
rather than multiple images due to strong gravitational lensing,
where these star-forming regions are unseen in the HST map due to the dust extinction and the large contrast to the bright emission from the accretion disk. 

We investigate whether the possibility of 1) is feasible with our fiducial best-fit mass model.  
In the middle and right panels of Figure \ref{fig:hst_crit_caus}, 
we present the caustics and the critical curve of our fiducial mass model  (Section \ref{sec:alma}). 
We find that our fiducial mass model can reproduce the morphology of the apparently single component in the HST map, 
if the position of the HST emission in the source plane is close to the cusp of the caustics. 
As an example, in Figure \ref{fig:hst_crit_caus} we show a possible position of the HST emission in the source plane (open blue triangle; middle panel) 
and the corresponding quadruple images in the image plane (filled blue triangle; right panel). 
We refer to the quadruple lens HST objects from bright to faint as HID1, HID2, HID3, and HID4. 
In this case, HID1, HID2, and HID3 are produced within a scale below the angular resolution of HST 
and are not resolved in the observed HST map. 
Since the sum of HID1, HID2, HID3 is over 100 times brighter than HID4, 
the HST emission appears as a single-like component in the observed HST map. 
The offsets among HID1, HID2, and HID3 can be 
even smaller 
if the position of the HST emission in the source plane is located closer to the cusp of the caustics. 
These results suggest that the possibility of 1) is indeed feasible. 
Note that the ALMA and HST emission in these cases have the offset of the $\sim50$ pc scale in the source plane that is reasonably smaller than the entire host-galaxy scale ($\sim$ kpc).
Interestingly, 
the spatial scale of $\sim$ 50 pc is consistent with the physical scale of a dense gas region around the central BH, called circumnuclear disk (CND; radius of $\sim100$ pc), 
that is formed by the inflowing gas due to the remaining angular momentum 
\citep[e.g.,][]{thompson2005,krips2007,hicks2009,davies2012,sani2012,izumi2013,izumi2015,garcia-burillo2014}. 
The intensive, dusty star-formation may occur in CND, 
cause the turbulence, and contribute to the angular momentum dissipation in CND, 
which helps the gas feeding on the accretion disk as well as 
the mass assembly of the SMBH \citep{kawakatu2009,izumi2016}. 

To quantitatively test the possibility of 1), 
we also carry out a simulation by making mock HST maps for 
different separations among HID1, HID2, and HID3. 
First, we identify isolated and unsaturated stars with SExtractor version 2.5.0 \citep{bertin1996} and evaluate the PSF of the HST map. 
The FWHM of the PSF is estimated to be $0.''103$ $\pm$ $0.''003$, where the central and error values are defined by the median and the standard deviation of the individual stars, respectively. 
Second, we make a mock HST map for the single-like component by injecting three PSFs at the positions of HID1, HID2, and HID3. 
Since the position of the HST emission in the source plane is sufficiently close to the cusp of the caustics, 
our fiducial mass model shows that the magnification factors among HID1, HID2, and HID3 maintain a $\sim$ two-one-one ratio
and that the positions of HID2 and HID3 are almost symmetry relative to HID1 along the line with the critical curve.  
We thus fix the 
brightnesses of these three PSFs at the two-one-one ratio and assume that the positions of HID2 and HID3 keep the same distance from HID1. 
Here we refer to the distance between HID1 and HID2 (= HID1 and HID3) as $l$. 
Finally, we measure the FWHM values for the single-like component in the mock HST map as a function of $l$.

%%%%%%%%%%%%%%%%%%%%%%%
\begin{figure}
\begin{center}
\includegraphics[trim=0cm 0cm 0cm 0cm, clip, angle=0,width=0.5\textwidth]{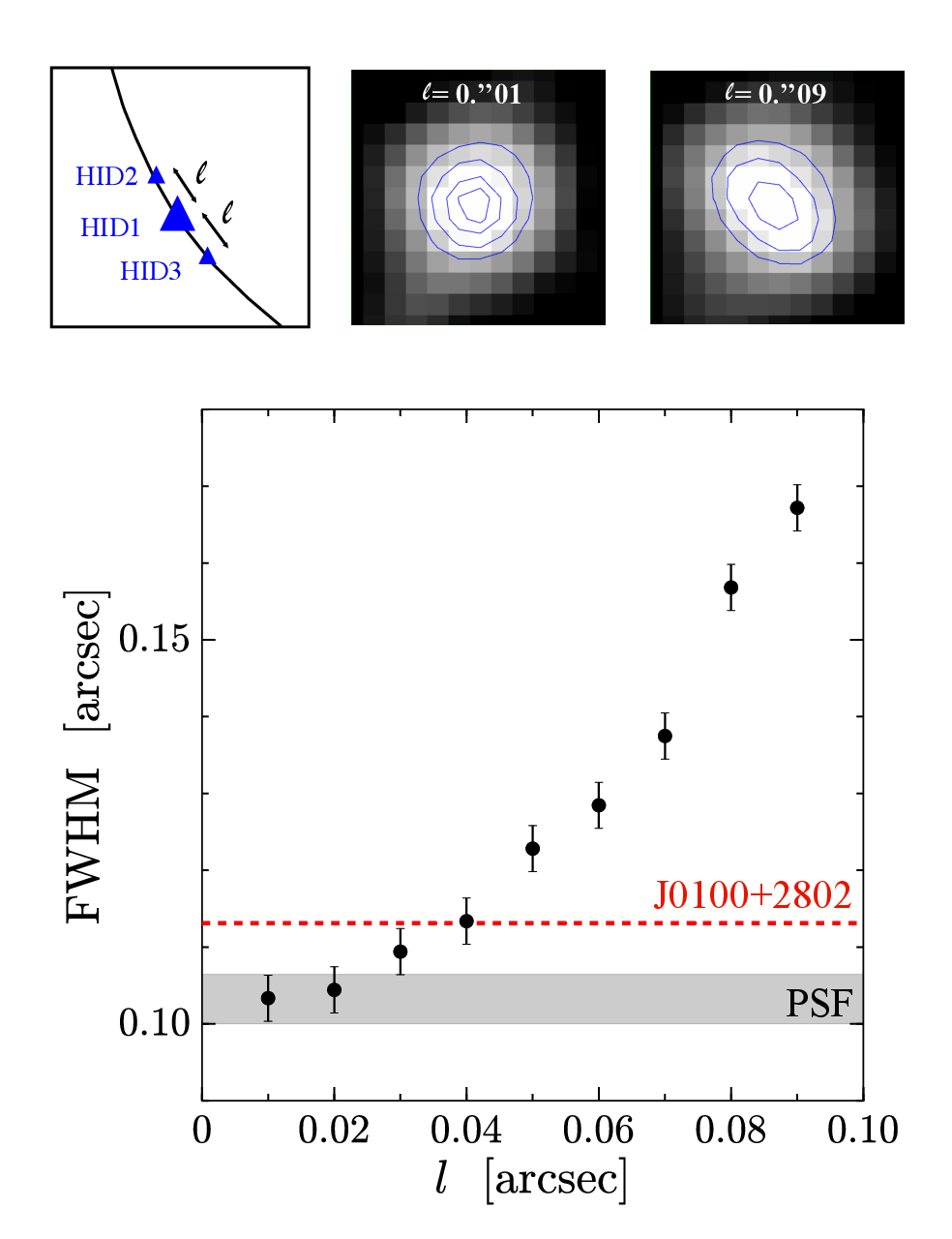}
 \caption[]{
Simulation results of the mock HST map for the single-like component. 
{\bf Top:}
Schematic overview of the simulations. 
The definition of $l$ and the mock HST maps in the cases of $l=0.''01$ and $l=0.''09$ 
are presented from left to right. 
{\bf Bottom:}
The black circles indicate the FWHM measurements as a function of $l$. 
The gray shaded region denotes the FWHM of the PSF estimated from the isolated and unsaturated stars, 
where the shade width is the uncertainty of the estimate. 
The red dashed line is the FWHM of J0100+2802 in the observed HST map. 
We use the uncertainty of the PSF as the error-bar scale of the black circles. 
\label{fig:off-fwhm}}
\end{center}
\end{figure}
%%%%%%%%%%%%%%%%%%%%%%% 

In Figure \ref{fig:off-fwhm}, we show the simulation results. 
The gray shaded region denotes the FWHM of the PSF including the uncertainty. 
For comparison, we also present the FWHM for J0100+2802 in the observed HST map that is estimated to be $0.''113$ (red dashed line).
We find that the FWHM of J0100+2802 exceeds that of the PSF, suggesting that the HST emission of J0100+2802 is not a point source. We also find that the FWHM of J0100+2802 corresponds to the simulation results with $l=0.''04$. 
In other words, the apparently single component in the HST map can be 
explained by the merging multiple images HID1, HID2, and HID3 with separation $l=0.''04$.  
These results indicate that the possibility of 1) 
is still consistent with the data.
In the case of $l=0.''04$, our fiducial mass model 
predicts the total magnification factor of  $\sim450$ for the quadruple
images in the HST map.

%%%%%%%%%%%%%%%%%%%%%%%%%%%%%%%%%%%%%%%%%%%%%%%%%%%%%%%%%%%%%%%%%
\section{Discussion}
\label{sec:discussion} 

The ALMA high-resolution map resolves the dust emission from J0100+2802 into four components whose positions and flux densities are explained by the gravitational lensing effect (Section \ref{sec:alma}). 
In contrast to the ALMA results, 
the high-resolution data of HST/F850LP shows the different morphology with 
an apparently single component,
but the detail morphological analysis with our fiducial lens mass model shows that it is possible to produce quadruple lens images that are consistent with the observed morphology in the HST map if we allow an offset between ALMA and HST emission regions (Section \ref{sec:hst}). 
Although we cannot draw definitive conclusion between the two possibilities of 1) the quadruply imaged lens system and 2) the multiple dusty star-forming system for J0100+2802 without further high-resolution observations,  in this section 
we focus on the possibility of 1) and discuss the intrinsic physical property of J0100+2802 in this scenario. 

In the case that the positions of HID1, HID2, and HID3 are very close with $l=0.''04$, 
our fiducial lens mass model estimates the total magnification factor of $\sim450$ for the quadruple images. 
The characteristic of J0100+2802 has been measured by the ground-based telescopes at the optical--NIR wavelengths \citep{wu2015} 
whose angular resolutions do not resolve the structure. Therefore, the previous measurements can be affected by the magnification factor of $\sim450$. 

\begin{figure}
\begin{center}
\includegraphics[trim=0cm 1.5cm 0cm 1.5cm, clip, angle=0,width=0.5\textwidth]{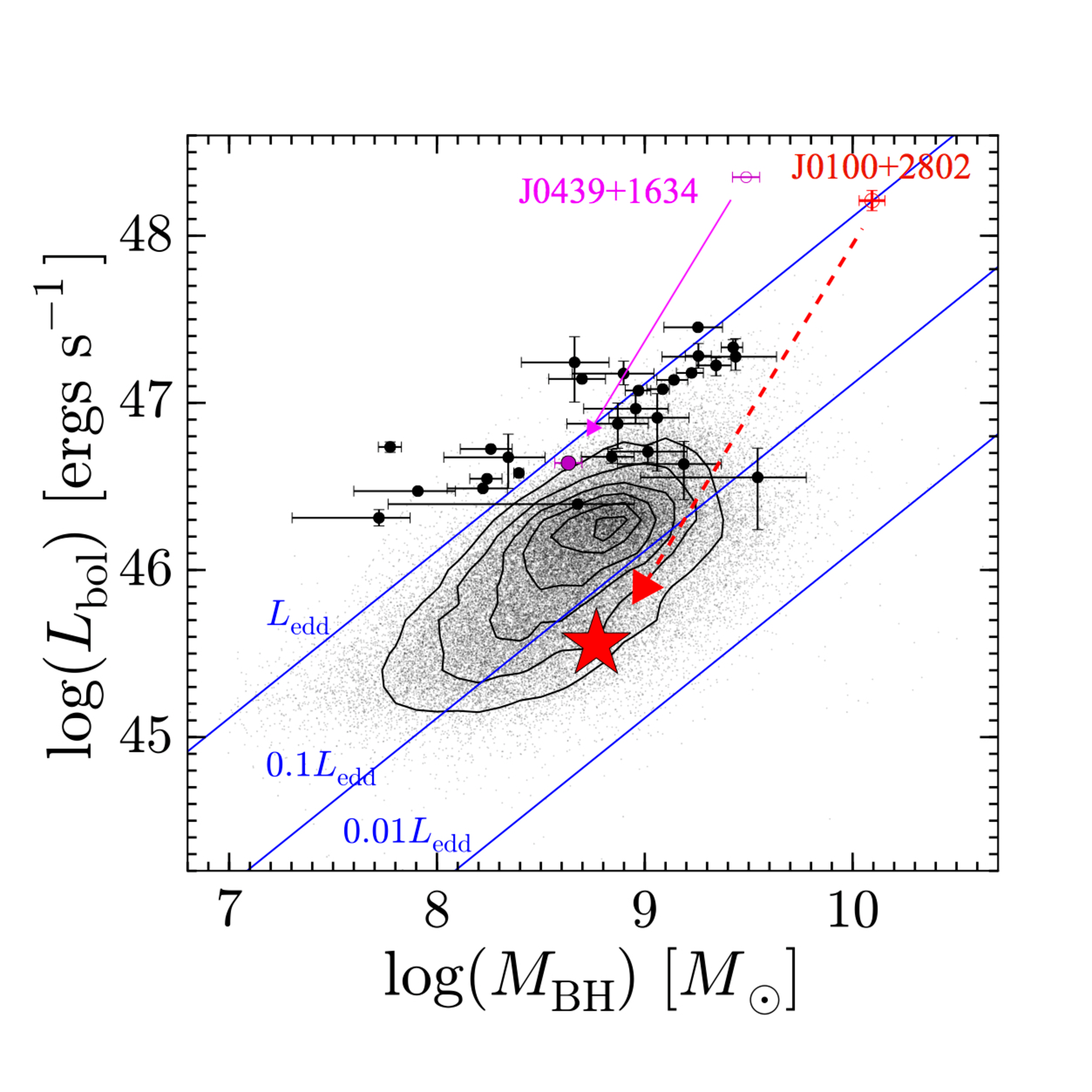}
 \caption[]{
Distribution of bolometric luminosity $L_{\rm bol}$ and BH mass $M_{\rm BH}$ estimated from Mg {\sc ii} line among QSOs. 
The red open circle indicates the apparent property of J0100+2802 \citep{wu2015}, 
while the red star denotes the potential intrinsic property after the gravitational lensing correct. 
For comparison, the magenta open and filled circles present the apparent and intrinsic properties of J0439+1634 that is identified as a gravitationally lensed QSO at $z=6.42$ \citep{fan2019}. 
The black circles are other QSOs at $z\gtrsim6$ \citep{mortlock2011,mazzucchelli2017,banados2018}, 
and the grey dots and black contours are the distribution of the SDSS QSOs at $z=0-2$ \citep{shen2011}. 
The blue lines present fractions of the Eddington luminosity. 
\label{fig:Mbh-Lbol}}
\end{center}
\end{figure}

The virial $M_{\rm BH}$ estimator based on the Mg {\sc ii} line \citep[e.g.,][]{vestergaard2009} is given by 
\begin{equation}
\label{eq:Mbh}
M_{\rm BH} = 10^{6.86} \, \biggl(\frac{\lambda L_{\lambda,3000}}{10^{44}{\rm erg\,s^{-1}}}\biggr)^{0.5} \,\biggl(\frac{\rm FWHM_{\rm MgII}}{\rm km\,s^{-1}}\biggr)^{2}, 
\end{equation}
where $\lambda L_{\lambda,3000}$ is the rest-frame UV luminosity at 3000 ${\rm \AA}$ wavelength and FWHM$_{\rm MgII}$ is the full-width-half-maximum of the Mg {\sc ii} line in the spectrum. 
If we apply the total magnification correction to the UV luminosity in Equation \ref{eq:Mbh}, 
the intrinsic $M_{\rm BH}$ estimate is decreased by a factor of $\sim$17.3. 
In Figure \ref{fig:Mbh-Lbol}, we show the intrinsic values of $L_{\rm bol}$ and $M_{\rm BH}$ of J0100+2802 after the total magnification correction. 
We find that J0100+2802 falls in the area within the distribution of low-redshift SDSS QSOs, 
and that the $M_{\rm BH}$ value is decreased even below $10^{9}\,M_{\odot}$. 
These results indicate that the gravitational lensing effect has a significant potential to change our understanding of the nature of the most massive SMBH at $z>6$. 
%%%%%%%%%%%%%%%%%%%%%%%
%%%%%%%%%%%%%%%%%%%%%%% 

We caution that our estimate of the total magnification factor of $\sim
450$ for the observed HST emission involves large uncertainty, given
that the lens mass model is not well constrained. Given the
discussions in e.g., \cite{keeton2003}, it is expected that the
total magnification near the cusp catastrophe for a given separation
of images in the image plane is roughly inversely proportional to the
ellipticity of the lens potential and also to $2-\alpha$, where
$\alpha$ denotes the radial slope of the lens potential ($\alpha=1$
for our fiducial SIE lens model). We find that the ellipticity is not
very well constrained in our model with $e=0.14\pm0.09$ ($1\sigma$),
which will lead to the uncertainty of the total magnification of a
factor $\sim 2$. Together with the fact that the radial slope may be
$\alpha\neq 1$, our estimate of the total magnification factor should
be uncertain by a factor of a few or so. 

Another caution is the proximity zone measurements of J0100+2802. 
The NIR spectroscopy towards J0100+2802 is reported to have a large proximity zone $R_{\rm p}$ of 7.8 $\pm$ 0.8 pMpc \citep{wu2015}. 
In \cite{eilers2017}, the authors carried out the deeper NIR spectroscopy with Keck/ESI, and updated the $R_{\rm p}$ measurement of $7.12 \pm 0.13$ pMpc. 
Taking the uniquely bright property into account, the authors find that this measurement is smaller than the prediction from the simulation ($\sim$ 12 pMpc) and discuss a possible scenario that this quasar has been shining for still shot such as $\sim10^{5}$ years. 
It is interesting that the gravitational lensing scenario is also consistent with the trend of the $R_{\rm p}$ measurement smaller than the prediction estimated from its luminosity. 

Note that we cannot rule out the possibility of 2).   
Recent ALMA observations reveal the overdensity of companion galaxies around the high-$z$ luminous QSOs \citep[e.g.,][]{decarli2017,trakhtenbrot2017}. 
The follow-up ALMA high-resolution observations also report the existence of the nearby on-going mergers around the $z>6$ QSOs \citep{venemans2019, banados2019}. 
The existence of the on-going mergers and/or the ISM environment harboring the multiple dusty star-forming regions in J0100+2802 will be definitely interesting and provide us with important insights to understand the rapid growth of the most massive SMBH in the early Universe. 
However, the Mg {\sc ii} absorption and the Ly$\alpha$ line features at $z=2.33$ in the optical--NIR spectroscopy implies the existence of the foreground galaxy that can causes gravitational lensing effect.
Indeed, the independent $M_{\rm star}$ estimates for the $z=2.33$ galaxy are consistent with each other (Section \ref{sec:alma}), 
although the uncertainties can be large due to the scaling relations. 
Moreover, recent HST observations report that another SDSS QSO at $z=6.34$, J0439+1634, known as the brightest QSO ($L_{\rm bol}=5.9\times10^{14}$ $L_{\odot}$) at $z>6$, is resolved into multiple objects with the existence of the nearby, low-luminous galaxy at the photometric redshift $z\sim0.7$ \citep{fan2019,pacucci2019}. 
The authors argue that the magnification factor is estimated to be $\sim$ 50 in their best-fit mass model that dramatically changes the intrinsic physical properties of J0439+1634 such as  
$L_{\rm bol}$ and $M_{\rm BH}$. 
Importantly, the $g$, $r$, $i$-band AB magnitudes of the foreground galaxy of J0439+1634 are $\sim$ 23--25 mag that are lower than the SDSS limiting magnitudes (5$\sigma\sim$ 21--22 mag) and thus negligible in the color diagnostics used for the QSO selection. 
Our and recent results suggest a possibility that a number of other bright QSOs at $z\gtrsim6$ are caused by the gravitational lensing effect by such low-luminous, but nearby foreground galaxies, 
even in the case that the morphology 
appears to consist of a single component.
We note that such large fraction of strong gravitational lensing among the apparently bright ($L_{\rm bol}\gtrsim10^{48}\,$ ergs s$^{-1}$) high-redshift QSOs are naturally explained by the magnification bias effect \citep{turner1980},
which tends to dominate at the bright end of any class of sources \citep[e.g.,][]{irwin1998,wyithe2002,negrello2010,quimby2014}.
\cite{pacucci2019b} calculate the servey volume and the probability of detecting one quasar with the magnification factor of $\sim$ 450 in the SDSS QSOs \citep{jiang2016}, finding that the slope in the intrinsic quasar luminosity function $\beta$ is required to be $\beta\geq3.7$. 
If this is true, the authors also argue that it is nearly impossible that the other remaining 51 SDSS QSOs at $z\gtrsim6$ are not magnified at least by the magnification factor of $\sim10$. 
The systematic high-resolution observations towards bright QSOs at $z>6$ are essential with ALMA, HST, and upcoming JWST. 

\section{Summary}
\label{sec:summary}
In this paper, we study the detailed morphology of the ALMA 1-mm continuum emission from the bright QSO at $z=6.30$, J0100+2802, 
that is known to contain the most massive BH at $z>6$ so far identified. 
In the high-resolution map, the emission is resolved into four components. 
We compare the ALMA results with the HST/F850LP data and investigate whether these four components are caused by the gravitational lensing effect,
which potentially has a significant impact on the estimate of intrinsic physical properties of J0100+2802.
The major findings of this paper are summarized below.
\begin{enumerate}
\item
We identify a clear continuum detection from J0100+2802 with the 21$\sigma$ level at the peak in the natural-weighted map, 
where the total flux density is estimated to be 1.3 mJy with the aperture diameter of $0\farcs8$. 
In the briggs-weighted map (robust = 0.2), the spatial resolution is improved about 40\% 
and the continuum emission is resolved into four components whose peak counts are all above the 5$\sigma$ level. 
\item 
We detect a possible Ly$\alpha$ line emission at the $5.3\sigma$ level at $z=2.33$ in the latest optical--NIR spectroscopy with VLT/Xshotter from the ESO archive, which is consistent with the Mg {\sc ii} absorption line feature at $z=2.33$ reported in the previous studies \citep{wu2015}. 
We construct a lens mass model assuming strong lensing by a galaxy at $z=2.33$ to find that a simple mass model  well reproduces the peak positions and flux densities for the four components identified in the ALMA map. 
\item 
The HST/F850LP map for J0100+2802 shows a morphology with an apparently single component, in marked contrast to the ALMA results. 
We discuss the following two possibilities for the difference in morphologies of J0100+2802 in the ALMA and HST maps: 
1) an intrinsic offset between the ALMA and HST emission in the source plane causes the different gravitational lensing effects and produces the different morphologies in the image plane. 
2) the four components in the ALMA map represent multiple dusty star-forming regions, rather than strongly lensed multiple images, where these star-forming regions are unseen in the HST map due to the dust extinction and the large contrast to the bright emission from the accretion disk. 
\item 
Assuming the possibility of 1), our fiducial best-fit lens mass model predicts the magnification factor of $\sim450$ for the observed HST emission.  
After the correction of gravitational lensing magnification, 
the intrinsic $L_{\rm bol}$ and $M_{\rm BH}$ relation falls in the area within the distribution of the low-redshift SDSS QSOs, 
and the $M_{\rm BH}$ measurement is decreased even below 10$^{9}\, M_{\odot}$.  
Our results suggest that the gravitational lensing effect has a significant potential to change our understanding of the most massive SMBH in the early Universe, and therefore should be explored further by future observations.
\end{enumerate}

We thank the anonymous referee for constructive comments and suggestions.
We are grateful to Nozomu Kawakatsu for organizing the productive workshop that made us start this study. 
We appreciate Shiro Mukae, Yoshiaki Ono, Ken Mawatari, Hiroshi Nagai, Yoichi Tamura, Akio Inoue, 
Xue-Bing Wu, Eduardo Ba$\tilde{\rm n}$ados, Zabl Johannes, Bo Milvang-Jensen, and Johan Peter Uldall Fynbo
 for providing us the helpful advice on analyzing the data. 
This paper makes use of ALMA data: ADS/JAO. ALMA \#2015.1.00692.S. 
ALMA is a partnership of the ESO (representing its member states), 
NSF (USA) and NINS (Japan), together with NRC (Canada), MOST and ASIAA (Taiwan), and KASI (Republic of Korea), 
in cooperation with the Republic of Chile. 
The Joint ALMA Observatory is operated by the ESO, AUI/NRAO, and NAOJ. 
This study is supported by World Premier International Research Center Initiative (WPI Initiative), 
MEXT, Japan, and KAKENHI (15H02064, 16J02344, 16K17670, 17H01114,17K14247, 18K03693)
Grant-in-Aid for Scientific Research (A) 
through Japan Society for the Promotion of Science (JSPS), 
Grant-in-Aid for Scientific Research (C) 
through JSPS,  Grant-in-Aid for Scientific Research on Innovative Areas,  
the Grant-in-Aid for JSPS Research Fellow, and the NAOJ ALMA Scientific Research Grant Number 2016-01A.
S.F. is supported by the ALMA Japan Research Grant of NAOJ Chile Observatory NAOJ-ALMA-231
and the Cosmic Dawn Center of Excellence funded by the Danish National Research Foundation under then grant No. 140. 

\bibliographystyle{apj}
\bibliography{apj-jour,reference}
%%%%%%%%%%%%%%%%%%%%%%%%%%%%%%%%%%%%%%%%%%%%%%%%%%%%%%%%%%%
\end{document}